\documentstyle[prl,aps,multicol,epsf]{revtex}
\input psfig
\begin {document}
\draft
\title {Velocity shift of surface acoustic waves due to interaction
with composite fermions in a modulated structure}
\author { A. D. Mirlin$^*$ and P. W\"olfle}
\address{Institut f\"ur Theorie der kondensierten Materie,
Universit\"at Karlsruhe, 76128 Karlsruhe, Germany}
\author {Y. Levinson}
\address {Department of
Condensed Matter Physics, The Weizmann Institute of Science,
Rehovot 76100, Israel}
\author{O. Entin-Wohlman}\address {School of Physics and Astronomy,
Raymond and Beverly Sackler Faculty of Exact Sciences,
\\ Tel Aviv University, Tel Aviv 69978, Israel}
\date {February 12, 1998}
\maketitle
\begin {abstract}
We study the effect of a periodic density modulation on surface
acoustic wave (SAW) propagation along a 2D electron gas 
near Landau level filling $\nu=1/2$. Within the composite
fermion theory, the problem is described in terms of fermions subject
to a  spatially modulated magnetic field and scattered by a random
magnetic field. We find that
a few percent modulation induces a large peak in the SAW velocity
shift, as has been observed recently by Willett {\it et
al}.  As further support of this theory we find the {\it dc}
resistivity to be in good agreement with recent data of Smet {\it et
al}.  
\end {abstract}
\begin{multicols}{2}
The interaction of  surface acoustic waves (SAW) with a 2D electron gas
(2DEG) in a strong magnetic field
was intensively studied during the last few years
\cite{Wix,Willett,Willett2}. It 
was found that the  velocity shift $\Delta v/v$ and the absorption
coefficient $\Gamma$ of the SAW show a rich structure reflecting
the magnetooscillations and Hall quantization of the
transport in a 2DEG. The SAW measurements provide a very
efficient method of studying the (frequency- and momentum-dependent)
conductivity of the 2DEG. The power of this method was demonstrated
in a study near half-filling ($\nu=1/2$) of the lowest
Landau level. It was found \cite{Willett2}
that the SAW velocity exhibits a pronounced
minimum near $\nu=1/2$, with a resonance structure appearing at 
very high SAW frequencies. These results are in good quantitative agreement
with theoretical calculations \cite{HLR,MW} based on the

Recently, new experimental results on SAW interaction with a 2DEG
 in the presence of a periodic array of parallel gates (grating)
 were obtained
\cite{WWP}. It was found that a weak modulation potential, 
with a wave vector
orthogonal to that of the SAW, changes the behavior of
$\Delta v/v$ near $\nu=1/2$ drastically, 
inducing a large peak similar to those
observed at quantum Hall states ($\nu=1,\,1/3$). 

The aim of this paper is to show that a large maximum in $\Delta v/v$ 
is precisely what the CF theory predicts in the
presence of a periodically modulated electric potential. 
In analogy to the Weiss
oscillation phenomenon in low magnetic fields \cite{WKPW89,B,we}, 
even a weak modulation produces a large 
correction to the effective conductivity, 
which is observed in the SAW measurements.

We consider the SAW (frequency $\omega$, wave vector ${\bf q}$, velocity
$v_s=\omega/q=2.8\times 10^5\,\mbox{cm/s}$ in GaAs) interacting with
a 2DEG of density $n$ in the presence of a periodic
potential (and, consequently, density) modulation with a wave vector
${\bf p}$. Near half-filling, the system is described \cite{HLR}
in terms of the CF's with Fermi wave vector $k_F=(4\pi n)^{1/2}$
moving in the reduced effective magnetic field
$B_{eff}=B-2hcn/e$. At low temperatures, 
the main source of CF scattering is the random
magnetic field due to the electron (or, equivalently, CF) density
inhomogeneity produced by the impurity
random potential \cite{HLR}. Similarly,
the main role of the grating will be in creating a magnetic field
modulation $\Delta B({\bf r})$ related to the density modulation as 
$\Delta B({\bf r})=-(2hc/e)\Delta n({\bf r})$, the direct effect of
the scalar potential modulation being smaller by a factor
$(p/k_F)^2\ll 1$. A general formalism for calculating the velocity
shift and the absorption coefficient in the presence of periodic
modulation has been recently developed in \cite{we}. The results are
expressed in the form
\begin{eqnarray}
&&\Delta v/v = (\alpha^2/2) \mbox{Re} (1+i\sigma_{eff}({\bf
q},\omega)/\sigma_m)^{-1}\ ; \label{dvv}\\
&&\Gamma = -q(\alpha^2/2) \mbox{Im} (1+i\sigma_{eff}({\bf
q},\omega)/\sigma_m)^{-1}\ , \label{gamma}
\end{eqnarray}
where $\alpha^2/2$ is the piezoelectric coupling constant,
$\sigma_m=\varepsilon v_s/2\pi$, and $\varepsilon$ is an effective
dielectric  constant of the background.
To present the expression for the effective conductivity
$\sigma_{eff}$, we introduce tensorial conductivities $\hat{
\sigma}^e_{ss'}$ with an index $s$ referring to the spatial Fourier
component with the wave vector ${\bf q}_s={\bf q}+s{\bf p}$ ($s=0,\pm
1,\pm 2,\ldots$), and the longitudinal conductivity $\sigma^e_{ss'}=
({\bf q}_s/q_s)\hat{\sigma}^e_{ss'} ({\bf q}_{s'}/q_{s'})$. 
The superscript $e$ serves
to distinguish the electronic conductivities from those of the CF's,
and the hat denotes the matrix structure in the $xy$ plane.
We will assume the density modulation to be of the
single-harmonic form, $\Delta n({\bf r})=\eta n\cos{\bf pr}$, 
where $\eta\ll 1$, and with a wave length much shorter than that of the
SAW, $p\gg q$. Then $\sigma_{eff}({\bf q},\omega)=\sigma^e(q,\omega)+
\delta\sigma_{eff}({\bf q},\omega)$, with
\begin{equation}
\delta\sigma_{eff}({\bf q},\omega)=
\sigma_{0,0}^{e(2)}
-\frac{\sigma_{0,1}^{e(1)}\sigma_{1,0}^{e(1)}+
\sigma_{0,-1}^{e(1)}\sigma_{-1,0}^{e(1)}}
{\sigma^e(p,\omega)-i(q/p)\sigma_{m}}.
\label{e1}
\end{equation}
Here the upper index $(i)$ refers to the $i$-th order of the expansion
in $\eta$; $\sigma^e(q,\omega)=\sigma_{0,0}^{e(0)}$ and 
$\sigma^e(p,\omega)=\sigma_{1,1}^{e(0)}$ being the longitudinal
conductivities at $\eta=0$. In the sequel, we will 
drop the superscript $(0)$, keeping $(1)$ and $(2)$ only. 

In the random phase approximation, 
the resistivity tensor of the electrons is related to that of the CF's
via \cite{HLR}
$\hat{\rho}_{ss'}^e=(2h/e^2)\hat{\epsilon}\delta_{ss'}+\hat{\rho}_{ss'}$,
where $\hat{\epsilon}$ is the antisymmetric tensor with
$\epsilon_{xy}=-\epsilon_{yx}=1$. This relation yields
\begin{eqnarray}
&&\hat{\sigma}_{00}^{e(2)} = -\hat{\sigma}_{00}^{e}\hat{\rho}_{00}^{(2)}
\hat{\sigma}_{00}^{e} + 2  
\hat{\sigma}_{00}^{e}\hat{\rho}_{01}^{(1)}
\hat{\sigma}_{11}^{e}\hat{\rho}_{10}^{(1)}
\hat{\sigma}_{00}^{e},   \label{e1a}\\
&&\hat{\sigma}_{01}^{e(1)} = -\hat{\sigma}_{00}^{e}\hat{\rho}_{01}^{(1)}
\hat{\sigma}_{11}^{e}, \qquad
\hat{\sigma}_{10}^{e(1)} = -\hat{\sigma}_{11}^{e}\hat{\rho}_{10}^{(1)}
\hat{\sigma}_{00}^{e},
\label{e1b}
\end{eqnarray}
where the grating-induced corrections to the resistivity tensor
can be related to the corrections to the CF conductivities 
$\hat{\sigma}_{ss'}^{(i)}$ by
\begin{eqnarray}
&& \hat{\rho}_{00}^{(2)}=
-\hat{\rho}_{00}\hat{\sigma}^{(2)}_{00}\hat{\rho}_{00}+
2\hat{\rho}_{00}\hat{\sigma}^{(1)}_{01}\hat{\rho}_{11}
\hat{\sigma}^{(1)}_{10}\hat{\rho}_{00},
\label{e2}\\
&&\hat{\rho}_{01}^{(1)}=
-\hat{\rho}_{00}\hat{\sigma}^{(1)}_{01}\hat{\rho}_{11}, \qquad
\hat{\rho}_{10}^{(1)}=
-\hat{\rho}_{11}\hat{\sigma}^{(1)}_{10}\hat{\rho}_{00}.
\label{e3}
\end{eqnarray}

To evaluate the 
CF conductivities, we use the
Boltzmann equation for the distribution function $F({\bf r},{\bf
n},t)$ in the presence of an 
electric field ${\bf E}({\bf r})={\bf E}_s\exp(i{\bf q}_s{\bf r})$,
\begin{eqnarray*}
&&\left\{-i\omega+v_F{\bf n\nabla}
+[\omega_c+\Delta\omega_c({\bf r})]
{\partial\over\partial\phi}-C\right\}F=v_Fe{\bf n}{\bf E}\ , 
\end{eqnarray*}
where ${\bf n}=(\cos\phi,\sin\phi)$ determines the direction of the
momentum, $\omega_c=eB_{eff}/mc$ is the cyclotron frequency, 
$\Delta\omega_c({\bf r})=e\Delta B({\bf r})/mc$, and $C$ is the
collision integral 
describing scattering by the random magnetic field with a transport
time $\tau$ \cite{MW}. To simplify the calculation, we assume the
low-momentum SAW field condition, $(ql)^2/(2\omega\tau)\ll 1$, where
$l$ is the CF mean free path (which
is marginally valid for $\omega=2\pi\cdot300\mbox{MHz}$, the lowest
frequency tested in the experiment \cite{WWP}). This will be
sufficient to explain this experiment, where no essential
dependence on $q$ was observed anyway. 
Under the condition assumed, we can set $q=0$ when calculating the
conductivities entering the r.h.s. of Eqs.~(\ref{e2}), (\ref{e3}). 
In particular,
$\hat{\sigma}_{00}$ is then approximated by the Drude form,
\begin{eqnarray}
&&\hat{\sigma}_{00}=\left(\hat{\rho}_{00}\right)^{-1}=
{\sigma_0(\tilde{\tau}/\tau)\over 1+\tilde{S}^2}\left(
\begin{array}{cc}          1       &      -\tilde{S}  \\
                        \tilde{S}  &       1
\end{array}
\right)\ ; \label{e6} \\
&&\sigma_0=(e^2/2h)k_F l\ ; \ \
\tilde{S}=\omega_c\tilde{\tau}\ ;\ \
\tilde{\tau}^{-1}=\tau^{-1}-i\omega\ .
\nonumber 
\end{eqnarray}
Furthermore, the grating-induced contributions
$\hat{\sigma}^{(2)}_{00}$, $\hat{\sigma}^{(1)}_{01}$ can be found via
the method developed in \cite{B,Gerh} as
\begin{eqnarray}
&& \hat{\rho}_{00}\hat{\sigma}^{(2)}_{00}\hat{\rho}_{00} =
{\eta^2\over 2}\left({2h\over e^2}\right)^2
\hat{\epsilon}\hat{\sigma}({\bf p},\omega)\hat{\epsilon}\ ;\nonumber \\
&&\hat{\rho}_{00}\hat{\sigma}^{(1)}_{01}=-\eta {h\over
e^2}\hat{\epsilon}\hat{\sigma}({\bf p},\omega);\ \   
\hat{\sigma}^{(1)}_{10}\hat{\rho}_{00} =-\eta {h\over e^2}
\hat{\sigma}({\bf p},\omega)\hat{\epsilon}.
\label{e7}
\end{eqnarray}

We choose 
${\bf p}\parallel {\bf e}_x$ and assume first
that the SAW wave vector (and the SAW electric field) is
orthogonal to ${\bf p}$, i.e. ${\bf q}\parallel {\bf e}_y$. 
Using Eq. (\ref{e7}), we find that the two terms in the expression for
$\hat{\rho}_{00}^{(2)}$, Eq. (\ref{e2}), cancel each other, 
while
$\hat{\rho}_{01}^{(1)}=\hat{\rho}_{10}^{(1)}=\eta(h/e^2)\hat{\epsilon}$. 
 Substituting  this into Eqs. (\ref{e1})--(\ref{e1b}) and using
$\omega_c\tau,\omega\tau\ll k_Fl$, we get
\begin{eqnarray}
\delta\sigma_{eff} & = & {\eta^2\over 2}\left[\sigma_{11,yy}^{e}-
{\sigma_{11,yx}^e\sigma_{11,xy}^e\over
\sigma_{11,xx}^e-i(q/p)\sigma_m}\right] \nonumber\\
& = & {\eta^2\over 2}
{1\over 
\rho_{11,yy}-i(q/p)(2h/e^2)^2\sigma_m}
\label{e8}
\end{eqnarray}
(we used also $(q/p)\sigma_m\rho_{11,xx}\ll 1$ in the second line).
This finally yields
$\sigma_{eff}$ in terms of the transverse resistivity of the CF's in the
absence of the grating,
\begin{eqnarray}
\sigma_{eff}({\bf q},\omega)&=&
(e^2/ 2h)^2\rho_{xx}({\bf q},\omega) \nonumber \\
&+&{\eta^2\over 2}{1\over\rho_{yy}({\bf
p},\omega)-i(q/p)(2h/e^2)^2\sigma_m}.
\label{e9}
\end{eqnarray}
The conductivity tensor $\sigma_{\mu\nu}({\bf p},\omega)$ 
was calculated in Ref. \cite{MW}, leading to the following result for the
transverse resistivity in terms of Bessel functions:
\begin{eqnarray}
&&\rho_{yy}({\bf p},\omega)=(iQS/ 4\sigma_0)[J_-/ J_{1-}-
J_+/ J_{1+}]; \label{e10} \\
&& J_{\pm}=J_{\pm [T+(1-2\beta)i/S]/(1\pm 2\beta i/S)}
(Q/(1\pm 2\beta i/S));\nonumber \\
&& J_{1\pm}=J_{1\pm [T+(1-2\beta)i/S]/(1\pm 2\beta i/S)}
(Q/(1\pm 2\beta i/S)), \nonumber
\end{eqnarray}
where $Q=pR_c$ with $R_c=v_F/\omega_c$ being the cyclotron radius,
$S=\omega_c\tau$, and $T=\omega/\omega_c$. Eq.~(\ref{e10}) is valid
for random magnetic field scattering ($\beta=1$), as well as for isotropic
potential scattering ($\beta=0$). 

Using Eqs.~(\ref{dvv}), (\ref{gamma}), 
(\ref{e9}), and (\ref{e10}), one can evaluate  the SAW
velocity shift and the absorption coefficient as functions of the
effective magnetic field. The velocity shift is plotted
 in Fig.~1 for the
experimentally relevant \cite{WWP} values of the density $n$, SAW
frequency  $\omega$, and the grating period
$d=2\pi/p$. 
We also used the typical experimental  values of the
CF effective mass $m=0.8m_e$ ($m_e$ being the free electron mass), the
CF transport relaxation time $\tau=40\mbox{ps}$, and the
conductivity parameter $\sigma_m=0.6\times 10^{6}\mbox{cm/s}$ \cite{note}.

As is seen from Fig.~1, a weak (3-5\%) modulation produces a large
peak in $\Delta v/v$, with an amplitude of the order of $1\times
10^{-4}$ (the scale is set by $\alpha^2/2=3.2\times 10^{-4}$ in GaAs),
i.e. approximately of the same magnitude as the maxima observed at the
quantum Hall fractions ($\nu=1,1/3$), in agreement with experiment
\cite{WWP,note2}. 
The reason for a weak modulation to be
sufficient to produce such a drastic effect is as follows. The
resistivities entering Eq.~(\ref{e9}) are of order of 
$\sigma_0^{-1}$, so that the second term becomes
comparable to the first one at $(\eta k_Fl)^2/2\sim 1$, which with
$k_F l\sim 50$ yields $\eta\sim 0.03$. Such an enhancement 
is familiar from the Weiss oscillations effect in low
magnetic fields \cite{WKPW89,B,we,Gerh}. 
For ${\bf q}\parallel{\bf p}\parallel {\bf e}_x$ all the indices $y$
are replaced by $x$ in the first line of 
(\ref{e8}), yielding a negligibly small
correction $\delta\sigma_{eff}$, in agreement with experiment.

\begin{figure}
\narrowtext
\centerline{\psfig{figure=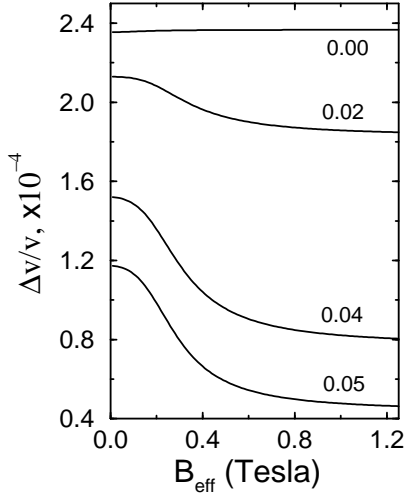,width=9cm}}
\vspace{-0.5cm}
\caption{Velocity shift $\Delta v/v$ of SAW as a function of the
effective magnetic field for
the following values of the parameters: $n=0.7\times
10^{11} {\rm cm}^{-2}$, $\omega=2\pi\cdot 300 {\rm MHz}$, $\tau=40 {\rm ps}$,
$d=0.7\mu{\rm m}$, $\sigma_m=0.6\times 10^6\mbox{cm/s}$. 
The strength $\eta$ of the grating  is shown on the curves.} 
\label{fig1}
\end{figure}

\vspace{-0.5cm}

\begin{figure}
\narrowtext
\centerline{\psfig{figure=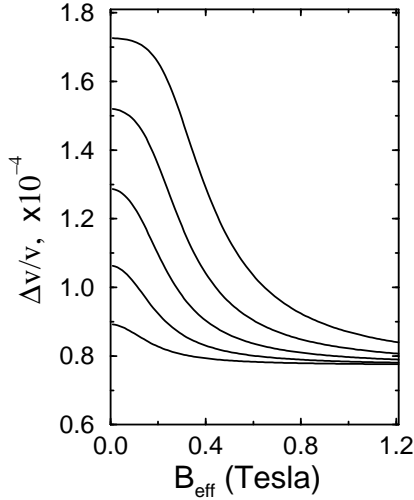,width=9cm}}
\vspace{-0.5cm}
\caption{Velocity shift for different values of the modulation period, 
from top to bottom: $0.5\mu{\rm m}$, $0.7\mu{\rm m}$,
$1.0\mu{\rm m}$, $1.5\mu{\rm m}$, $2.5\mu{\rm m}$. The modulation
strength is $\eta=0.04$ for all the curves.}
\label{fig3}
\end{figure}

In the experiment \cite{WWP}, the peak was observable only for short
enough grating period, $d=2\pi/p\le 1.5\mu\mbox{m}$. To
illustrate the 
theoretical dependence of the magnitude of the peak on $d$, 
we plot in Fig.~2 the results for  $\Delta v/v$
for a fixed grating amplitude and different values of $d$. It is
seen that indeed the peak height becomes rather small for $d\gtrsim
2\mu\mbox{m}$ (which corresponds to a grating period
considerably larger than the mean free path $l$, equal to
$0.7\mu\mbox{m}$ for the parameters used). 

We comment now on the role played by the scattering mechanism. In
Fig.~3 the SAW velocity shift is plotted for scattering by a random
magnetic field and
by a random potential, respectively, using the same value of the transport
time $\tau$ and keeping all other parameters  
fixed. While the amplitude of the peak is approximately
the same in both cases, the
shape is very different. For random potential scattering, the peak
is considerably sharper, with an oscillatory structure, not
observed in \cite{WWP}. In contrast, for random magnetic field
scattering, we find a broad and smooth peak, the shape and width of which
are in agreement with the experiment \cite{WWP}. 

\begin{figure}
\narrowtext
\centerline{\psfig{figure=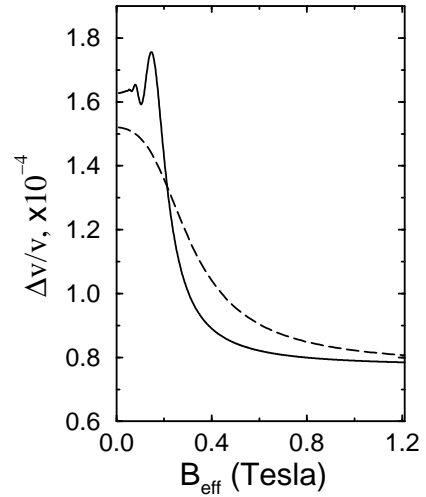,width=9cm}}
\vspace{-0.5cm}
\caption{Velocity shift for the random magnetic field (dashed line) and
random potential (full line) scattering at the same values of 
the parameters ($\eta=0.04$, $d=0.7\mu\mbox{m}$).}
\label{fig4}
\end{figure}

There are some differences between the experimental and theoretical
results, for which the present theory does not seem to account.
The theoretical value of $\Delta v/v$ in the center of the peak
decreases with $\eta$, while it increased in the experiment.  
Also, the theory
predicts that the peak width increases with $p$ (at $p\gg q$), while
in the experiment the width was weakly dependent on $p$ and
$q$. We think that the latter feature does not have a deep 
meaning and may only be valid in a restricted range of the parameters. 

Finally, the grating-induced correction to the {\it dc}
conductivity is given by the first term in Eq.~(\ref{e1}),
$\sigma_{00}^{e(2)}$ (in the limit $q\to 0$ and then $\omega\to 0$),
which yields, according to the above calculation,
\begin{equation}
\hat{\sigma}_{00}^{e(2)}=(\eta^2/ 2)\hat{\sigma}({\bf p}, 0).
\label{dc1}
\end{equation}
Since $\sigma_{yy}({\bf p},0)$ is the only
non-zero component of $\hat{\sigma}({\bf p},0)$,
Eq. (\ref{dc1}) implies a correction to the $xx$-component of the
macroscopic resistivity tensor
$\hat{\rho}_{dc}=(\hat{\sigma}_{00}^e)^{-1}$ 
measured in {\it dc} experiments \cite{note3} 
\begin{equation}
\rho_{dc,xx}=\sigma_0^{-1}+2\eta^2(h/e^2)^2\sigma_{yy}({\bf p},0)
\label{dc2}
\end{equation}
An experiment on {\it dc} transport near $\nu=1/2$ in a modulated
structure was performed recently by Smet et al  \cite{Smet}. In Fig.~4 we
present the theoretical result, Eq. (\ref{dc2}), for the parameters of
the sample A (Fig.~1 of \cite{Smet}). The modulation amplitude is
estimated from the comparison with experiment as $\eta=0.026$, which
is in good agreement with $\eta_l=0.032$ found from the fit of the
low-field Weiss oscillations 
(also presented in Fig.~1 of \cite{Smet}) to the
theoretical formula \cite{MW1} taking into account 
the scattering by the smooth random potential.
As is seen from Fig.~4, the theoretical results reproduce 
well the width of the grating induced minimum in $\rho_{dc,xx}$ around
$\nu=1/2$. The commensurability oscillations are again (almost) washed
out due to the scattering by the random magnetic field. Only the first
minimum is 
weakly developed at $B_{eff}\approx 0.4 T$, possibly corresponding to
the shoulders observed experimentally. The difference between the
theoretical and the experimental curve at small $B_{eff}$ (V-shaped
minimum vs. plateau) is probably due to open CF orbits \cite{Beton},
which are not taken into account
by our perturbative-in-$\eta$ calculation and
should produce an additional positive magnetoresistance
in a range $|B_{eff}|<B_c=\eta(2hcn/e)\approx 0.4 T$.

\begin{figure}
\narrowtext
\centerline{\psfig{figure=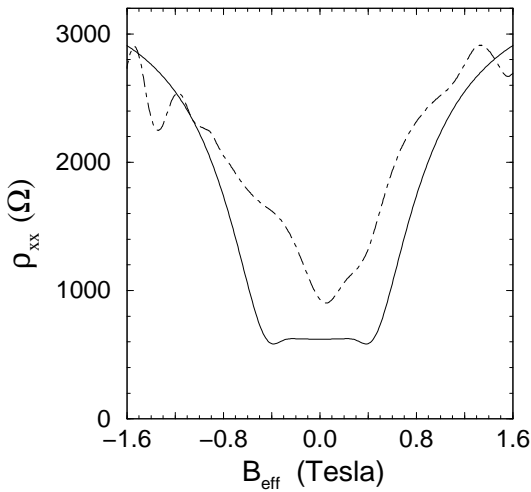,width=7cm}}
\caption{{\it dc} magnetoresistivity. Dash-dotted
line -- experiment of Ref.~\protect\cite{Smet}, solid line -- theory
for the experimental parameters $d=400 \mbox{nm}$, 
$n=1.8\times 10^{11}\mbox{cm}^{-2}$, $\sigma_0^{-1}=270\Omega$, and
with $\eta=0.026$. } 
\label{fig5}
\end{figure}
\vspace{-0.3cm}

In conclusion, we have studied the propagation of SAW interacting with
a 2DEG near $\nu=1/2$ in the presence of a weak density
modulation. Within the CF theory, the problem is described in terms of
fermions subject to a
 modulated magnetic field and scattered by a random magnetic
field. Using the Boltzmann equation approach, we have calculated the
SAW velocity shift and found that it exhibits, at modulation strength
$\sim 3\%$, a pronounced maximum, with amplitude of order
of the piezoelectric coupling constant, in agreement with the
experiment \cite{WWP}. The calculated correction to the {\it dc}
resistivity describes reasonably well the magnitude and the width of
the minimum of $\rho_{xx}$ near $\nu=1/2$ observed in \cite{Smet}.

While working on this project, we became aware of the preprint
\cite{OSH}, where the same problem was addressed. 
The authors of \cite{OSH} arrived at a formula similar to our
Eq.~(\ref{e9}), and then proceeded via numerical solution of the
Boltzmann equation for isotropic scattering. 
Our result for the
{\it dc} case, Eq.~(\ref{dc2}), is however different from that
obtained in \cite{OSH}. 

We thank A.~Stern, F.~von~Oppen, J.H.~Smet, K.~von~Klitzing, and
 R.L.~Willett  for discussions of unpublished results, and J.H.~Smet
 for sending us the experimental data of Ref.~\cite{Smet}. 
This work was supported by the Alexander von Humboldt Foundation
(Y.~L.), the Albert Einstein Minerva Center for Theoretical Physics
(O.~E.-W.), the SFB195 der Deutschen
Forschungsgemeinschaft and the German--Israeli
Foundation  (A.~D.~M. and P.~W.).

\end{multicols}

\begin{references}
\bibitem{note2} Note that for the relatively low frequency assumed in
Fig.~1 the minimum in $\Delta v/v$ at $\eta=0$ is very weak.
\end{references}
\end {document}